\documentclass[11pt,twoside]{article} 
\usepackage{asp2004}
\usepackage{epsf}
\usepackage{psfig}
\usepackage{lscape} 

\begin{document} 
\title{The Spectral Variability of Pulsating Stars: PG\,1159-035}
\author{T. Stahn, S. Dreizler} 
\affil{Institut f\"ur Astrophysik, Universit\"at G\"ottingen,
  Friedrich-Hund-Platz 1, D-37077 G\"ottingen, Germany}
\author{K. Werner} 
\affil{Institut f\"ur Astronomie und Astrophysik, Universit\"at
  T\"ubingen, Sand 1, D-72076 T\"ubingen, Germany}
%
%
\begin{abstract} 
With 10m class telescopes as well as with time-tagging detectors on
board of HST and FUSE, the analysis of time-resolved spectra for
pulsating white dwarfs becomes feasible. We present simulated time-resolved
spectra for the hot pulsating white dwarf PG\,1159-035 and compare
these models with observational data of the 516\,s mode based on
HST-STIS spectroscopy. A determination of the pulsation mode by the
spectral variability of PG\,1159-035 seems to be impossible for the moment.
\end{abstract}
%
%
\section{Introduction}
Analyses of stellar oscillations are important tools for probing
stellar structure and evolution and hence for analyses of white dwarfs
as well. Variable white dwarfs are non-radial g-mode-pulsators whose
pulsation modes can be described in terms of spherical harmonics with
parameters $l$ and $m$. PG\,1159-035 is the prototype for hot
pulsating H-deficient white dwarf stars at the transition stage to the
white dwarf sequence. The response of stellar atmospheres to
pulsations can be seen in wavelength-dependent flux variations. We
calculate these wavelength-dependent flux variations for modes with
different $l$ and compare them with observational data based on
HST-STIS spectra. The goal is to get an independent method for
analyses of pulsation properties and the determination of pulsation
modes as a supplement to photometric analyses
\citep[e.g.][]{1991ApJ...378..326W,1994ApJ...427..415K}.
\par
In this paper, we describe our first approach towards the spectral
analysis of the pulsating white dwarf PG\,1159-035. We begin in
Section 2 with a description of the model properties on which our
analysis is based. In Section 3, we present calculations of chromatic
amplitudes for different pulsation modes and compare these with
observational data in Section 4. We present our conclusions in \mbox{Section
5.}
%
%
\section{Spectral Analysis of PG\,1159-035}\label{model_atmos}

Our analysis is based on theoretical spectra which are calculated with
a NLTE model atmosphere program \citep{werner_dreizler_1999,2003sam..conf...69D,2003sam..conf...31W} using detailed
\mbox{H-He-C-O} model atoms \citep{2004A&A...421.1169W}. Our model grid
ranges from \mbox{T$_{\rm eff}$=137000-143000\,K} (in steps of 250\,K)
for \mbox{$\log{g}=7.0$} and \mbox{$\log{g}=6.8-7.2$} (in steps of
0.1\,dex) for \mbox{T$_{\rm eff}$=140000\,K}, respectively. These
values are distributed around the parameters \mbox{T$_{\rm
eff}$=140000\,K} and \mbox{$\log{g}=7.0$} derived in earlier analyses
\citep{1991A&A...244..437W,1998A&A...334..618D}.  
\par
\begin{figure}
\plottwo{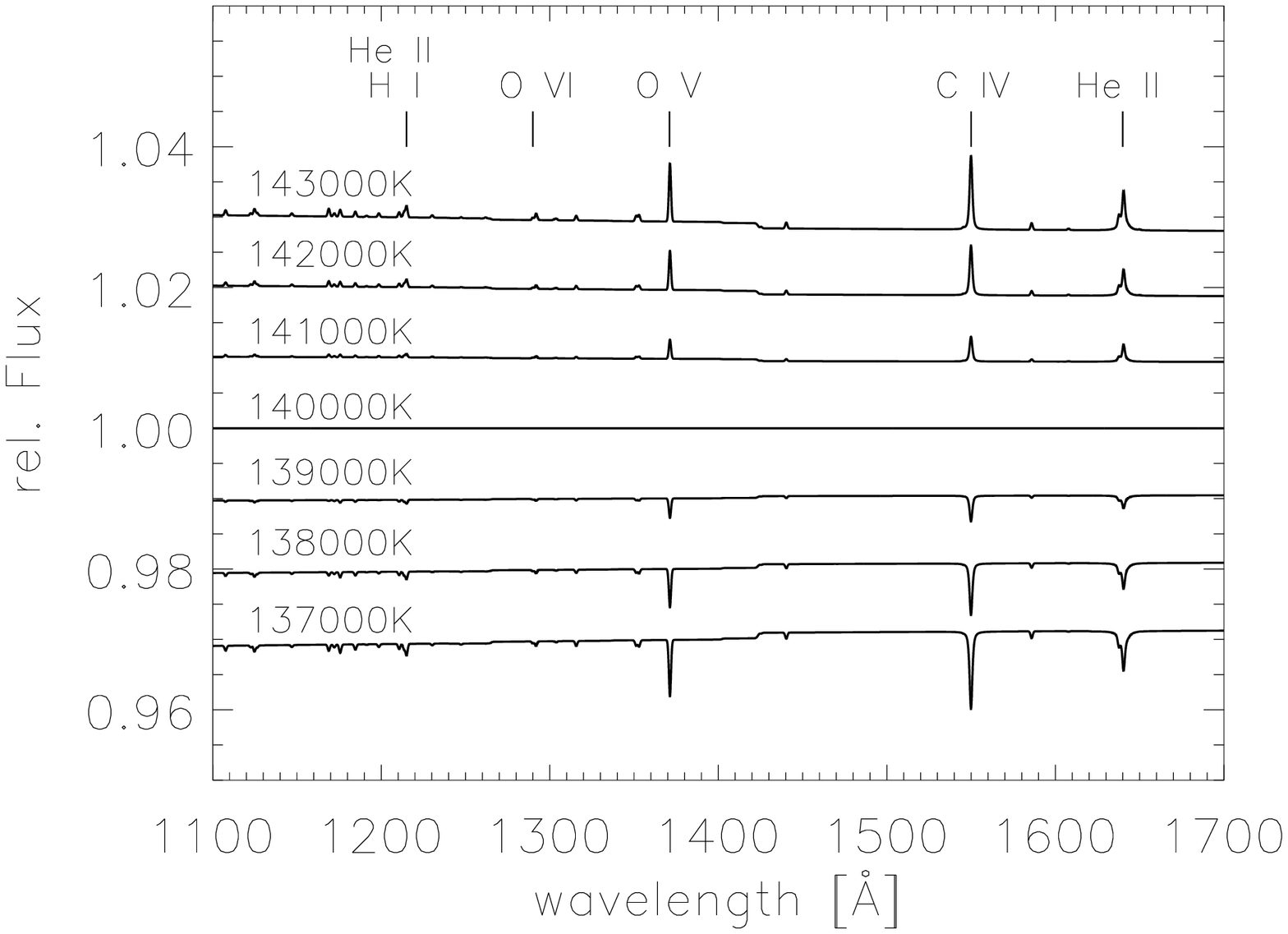}{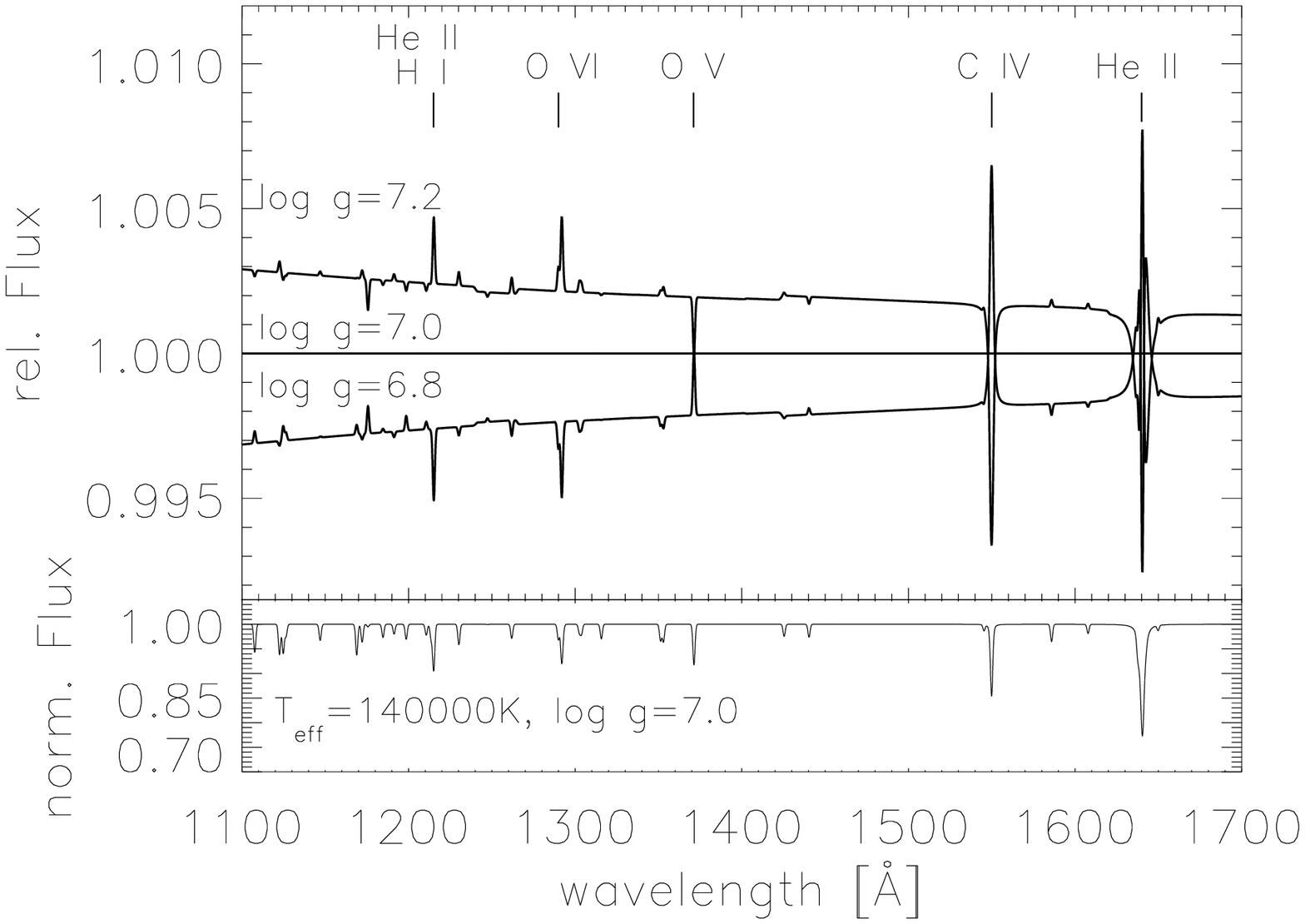}
\caption{Flux variation relative to the best fit model with T$_{\rm
    eff}$=140\,kK and $\log{g}=7.0$. The left plot shows flux
    variations for several temperatures and fixed $\log{g}=7.0$. The
    right plot shows flux variations for several values of $\log{g}$
    and a fixed temperature of T$_{\rm eff}$=140\,kK (upper panel) as
    well as a model of the normalized flux for T$_{\rm eff}$=140\,kK
    and $\log{g}=7.0$ (lower panel). All spectra are convolved with a
    1.5\,\AA\ Gaussian.}
\label{spectrum}
\end{figure}

 
Figure \ref{spectrum} shows a normalized model spectrum for T$_{\rm
eff}$=140\,kK and $\log{g}=7.0$ as well as flux variations relative to
this model caused by variations in temperature and surface gravity,
respectively.
%
%
\section{Chromatic Amplitudes}\label{chrom_amp}

According to \citet{1982ApJ...259..219R,1995ApJ...438..908R} we calculate the
wavelength-dependent flux variations between 200\,\AA\ and 7000\,\AA\ by
\begin{eqnarray*}
  F_{\lambda}&\propto& I_{0\lambda} \int{h_{\lambda}(\mu)\;\mu
    \;d\mu}\\
  \Delta F_{\lambda} &\propto& I_{0\lambda} \left(
  \frac{1}{I_{0\lambda}} \frac{\partial I_{0\lambda}}{\partial T}
  \right)  \int{h_{\lambda}(\mu)\; P_l(\mu) \;\mu
    \;d\mu}\nonumber
\end{eqnarray*}
where $P_l(\mu)$ is the Legendre polynomial of \mbox{degree $l$}
(all models with $m$=0) and the integrals are evaluated
between $\mu =0$ and $\mu =1$. We derive a limb darkening
$h_{\lambda}$ from nine angle values of $\mu$ between $\mu=0$ and
$\mu=1$ for each wavelength in each spectrum and replace the
derivative $(\partial I_{0\lambda}/ \partial T)$ by a difference
quotient $\Delta I_{0\lambda}/\Delta T$ computed with models of
$T_{\rm eff}$=139750\,K and $T_{\rm eff}$=140250\,K for each
wavelength.
\par
\begin{figure}[!ht]
\plotfiddle{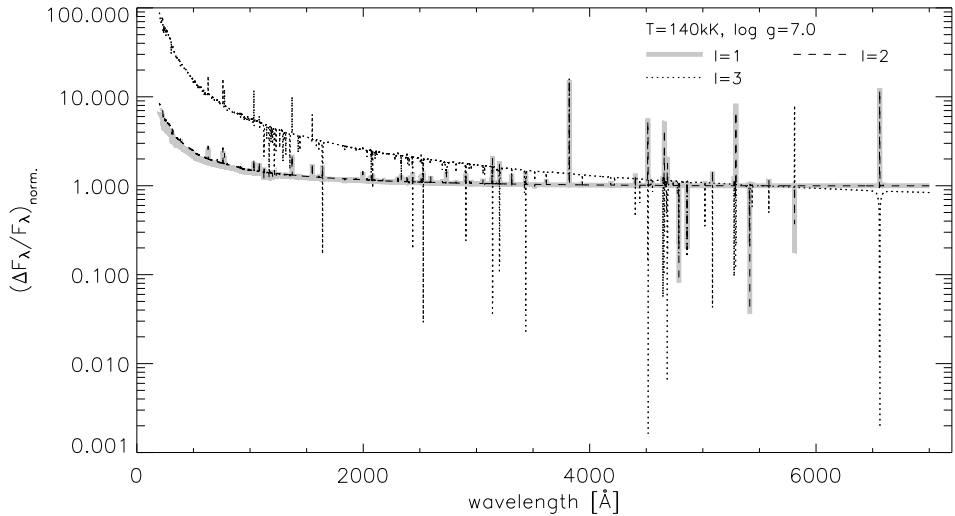}{6.5cm}{0}{50}{50}{-210}{-180}
\caption{Chromatic amplitudes $\Delta F_{\lambda}/F_{\lambda}$
    calculated for T$_{\rm eff}$=140000\,K and $\log{g}=7.0$ and for
    pulsation modes with $l$=1-3 and wavelengths ranging from
    200-7000\,\AA; chromatic amplitudes are normalized at 5500\,\AA\
    and are convolved with a 1.5\,\AA\ Gaussian.}
\label{chromatic_amplitude}
\end{figure}

Figure \ref{chromatic_amplitude} shows our results of calculated
chromatic amplitudes $\Delta F_{\lambda}/F_{\lambda}$ using the best
model with parameters T$_{\rm eff}$=140000\,K, $\log{g}=7.0$ for
pulsation modes with $l$=1-3 ($m$=0).
\par
Unfortunately, our models show almost no difference between
pulsation modes with $l$=1 and $l$=2 in nearly the whole wavelength
range, especially for wavelengths between 1100\,\AA\, and  \mbox{1800\,\AA}. 
%
%
\section{Observation}\label{observ}      

\begin{figure}[!ht]
\plotfiddle{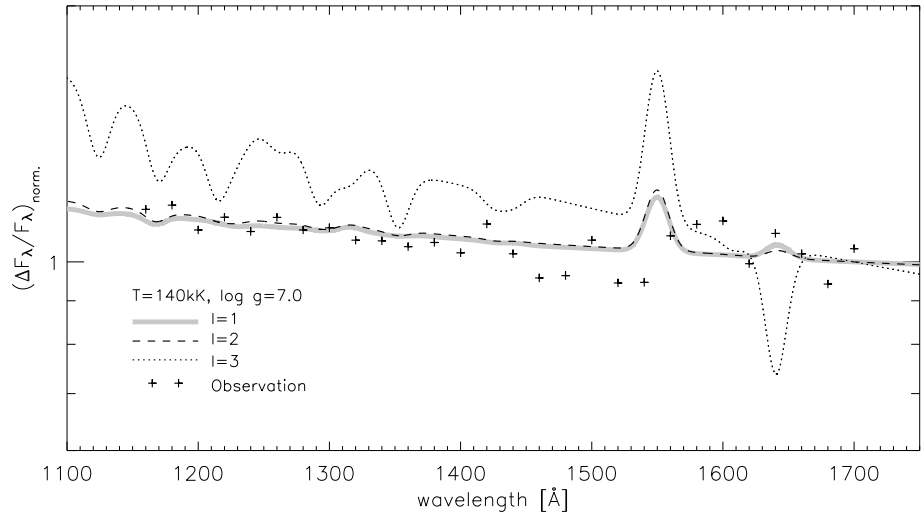}{6.5cm}{0}{50}{50}{-225}{-180}
\caption{Chromatic amplitudes as in Figure \ref{chromatic_amplitude}
    compared with the 516\,s mode of PG\,1159-035 obtained from
    HST-STIS spectroscopy. The models are convolved with a 20\,\AA\
    Gaussian in order to match the observational binning.}
\label{observation}
\end{figure}

The observational time-resolved spectra of PG\,1159-035 were obtained
during Cycle 10 with the HST-STIS. The observations were obtained with
the G140L grating and a 52x0.2 slit in time-tagged mode. The 20 orbits
are grouped into seven blocks of about 40 min exposure
time. Afterwards, the 20 orbits are split into 955 blocks of about
50\,s during data reduction with IRAF SDSDAS routines. The observed
chromatic amplitudes were obtained by binning the spectra by 20\,\AA\,
and fitting sine functions with periods fixed to the modes determined
from WET observations \citep{1991ApJ...378..326W}. Observed chromatic
amplitudes shown in Figure \ref{observation} are obtained from the
most prominent pulsation mode with a period of P=516.040\,s. For
comparison, Figure \ref{observation} shows our calculated models for
pulsation modes with $l$=1-3, as well. Models are normalized at
1700\,\AA\ and the observations are normalized at the mean value of
chromatic amplitudes for \mbox{1660\,\AA}, \mbox{1680\,\AA}, and
1700\,\AA\ in order to remove systematic deviations between the models
and the observational data.
\par
Figure \ref{observation} doesn't allow the determination of the
pulsation mode of the 516\,s period: although the observation roughly
matches the slope of the $l$=1 and $l$=2 mode, the difference between
these models are too small and the fluctuation of the observational
data around the models is too high for further analyses.
%
%
\section{Summary and Conclusions}\label{conclusion}

We have calculated the chromatic amplitudes for pulsation modes with
$l$=1-3 to analyze the spectral variability of the hot pulsating white
dwarf PG\,1159-035. Our models show almost no difference
between modes with $l$=1 and $l$=2 in the UV and optical wavelength
range. For the moment therefore, it seems to be impossible to
determine pulsation modes of GW Vir pulsators by their spectral
variability.
\par
There are two obvious means of improving our results: we can expand
the wavelength range of our observational data to shorter wavelengths,
e.g. with FUSE data; and - from a theoretical point of view - we will
try to improve our models by surface-resolved flux synthesis.
%
%
\acknowledgements{T. Stahn would like to thank the organizers of the
    workshop for financial support.
    \par
    Observations were made with NASA/ESA Hubble Space Telescope,
    obtained from the data archive at the Space Telescope Institute.}
    \par
%
%
%

\begin{thebibliography}{}
\expandafter\ifx\csname natexlab\endcsname\relax\def\natexlab#1{#1}\fi

\bibitem[{{Dreizler}(2003)}]{2003sam..conf...69D}
{Dreizler}, S. 2003, in ASP Conf. Ser. 288: Stellar Atmosphere Modeling, ed.
  I.~Hubeny, D.~Mihalas, \& K.Werner, 69

\bibitem[{{Dreizler} \& {Heber}(1998)}]{1998A&A...334..618D}
{Dreizler}, S. \& {Heber}, U. 1998, \aap, 334, 618

\bibitem[{{Kawaler} \& {Bradley}(1994)}]{1994ApJ...427..415K}
{Kawaler}, S.~D. \& {Bradley}, P.~A. 1994, \apj, 427, 415

\bibitem[{{Robinson} {et~al.}(1982){Robinson}, {Kepler}, \&
  {Nather}}]{1982ApJ...259..219R}
{Robinson}, E.~L., {Kepler}, S.~O., \& {Nather}, R.~E. 1982, \apj, 259, 219

\bibitem[{{Robinson} {et~al.}(1995){Robinson}, {Mailloux}, {Zhang}, {Koester},
  {Stiening}, {Bless}, {Percival}, {Taylor}, \& {van
  Citters}}]{1995ApJ...438..908R}
{Robinson}, E.~L., {Mailloux}, T.~M., {Zhang}, E., {et~al.} 1995, \apj, 438,
  908

\bibitem[{Werner \& Dreizler(1999)}]{werner_dreizler_1999}
Werner, K. \& Dreizler, S. 1999, in Computational Astrophysics, ed. H. Riffert
  \& K. Werner, Journal of Computational and Applied Mathematics, Vol. 109, 65

\bibitem[{{Werner} {et~al.}(1991){Werner}, {Heber}, \&
  {Hunger}}]{1991A&A...244..437W}
{Werner}, K., {Heber}, U., \& {Hunger}, K. 1991, \aap, 244, 437

\bibitem[{{Werner} {et~al.}(2003){Werner}, {Deetjen}, {Dreizler}, {Nagel},
  {Rauch}, \& {Schuh}}]{2003sam..conf...31W}
{Werner}, K., {Deetjen}, J.~L., {Dreizler}, S., {et~al.} 2003, in ASP Conf.
  Ser. 288: Stellar Atmosphere Modeling, ed. I.~Hubeny, D.~Mihalas, \&
  K.Werner, 31

\bibitem[{{Werner} {et~al.}(2004){Werner}, {Rauch}, {Barstow}, \&
  {Kruk}}]{2004A&A...421.1169W}
{Werner}, K., {Rauch}, T., {Barstow}, M.~A., \& {Kruk}, J.~W. 2004, \aap, 421,
  1169

\bibitem[{{Winget} {et~al.}(1991){Winget}, {Nather}, {Clemens}, {Provencal},
  {Kleinman}, {Bradley}, {Wood}, {Claver}, {Frueh}, {Grauer}, {Hine}, {Hansen},
  {Fontaine}, {Achilleos}, {Wickramasinghe}, {Marar}, {Seetha}, {Ashoka},
  {O'Donoghue}, {Warner}, {Kurtz}, {Buckley}, {Brickhill}, {Vauclair}, {Dolez},
  {Chevreton}, {Barstow}, {Solheim}, {Kanaan}, {Kepler}, {Henry}, \&
  {Kawaler}}]{1991ApJ...378..326W}
{Winget}, D.~E., {Nather}, R.~E., {Clemens}, J.~C., {et~al.} 1991, \apj, 378,
  326

\end{thebibliography}

\end{document}